 \documentclass[twocolumn]{aastex631}

\shorttitle{Light Curve Analysis of YZ Ret}
\shortauthors{Hachisu \& Kato}

\begin{document}


\title{Light-curve analysis of the classical nova YZ Ret: revisit
with fully self-consistent nova outburst models}


\author[0000-0002-0884-7404]{Izumi Hachisu}
\affil{Department of Earth Science and Astronomy, 
College of Arts and Sciences, The University of Tokyo,
3-8-1 Komaba, Meguro-ku, Tokyo 153-8902, Japan} 
\email{izumi.hachisu@outlook.jp}




\author[0000-0002-8522-8033]{Mariko Kato}
\affil{Department of Astronomy, Keio University, 
Hiyoshi, Kouhoku-ku, Yokohama 223-8521, Japan} 

%
%




\begin{abstract}
YZ Ret is a well observed nova from the first detection of an X-ray flash
phase in classical novae until the end of a supersoft X-ray source phase. 
We have reanalyzed multiwavelength light curves of YZ Ret over 
the total duration of the outburst.  
Our fully self-consistent nova outburst model, 
a 1.25 $M_\sun$ white dwarf (WD) and a mass accretion rate 
of $1\times 10^{-9} ~M_\sun$ yr$^{-1}$ on to the WD,  
reasonably reproduced $V$/$g$ light curves as well as the very early
X-ray flash and late supersoft X-ray light curves.
This model explains the emergence epoch of the GeV gamma-ray emission
because a shock naturally arises far outside the nova (WD) photosphere 
just after the optical maximum.  In our model, before optical maximum,
later ejected matter has a smaller velocity than that of earlier ejecta
and therefore the ejecta expands while, after optical maximum,
later ejected matter has a larger velocity than that of earlier ejecta
and collision with the earlier ejecta makes a shock.
\end{abstract}


\keywords{gamma-rays: stars --- novae, cataclysmic variables --- 
stars: individual (YZ~Ret) --- stars: winds --- X-rays: stars}


\section{Introduction}
\label{introduction}

\begin{figure*}
\epsscale{0.75}
\plotone{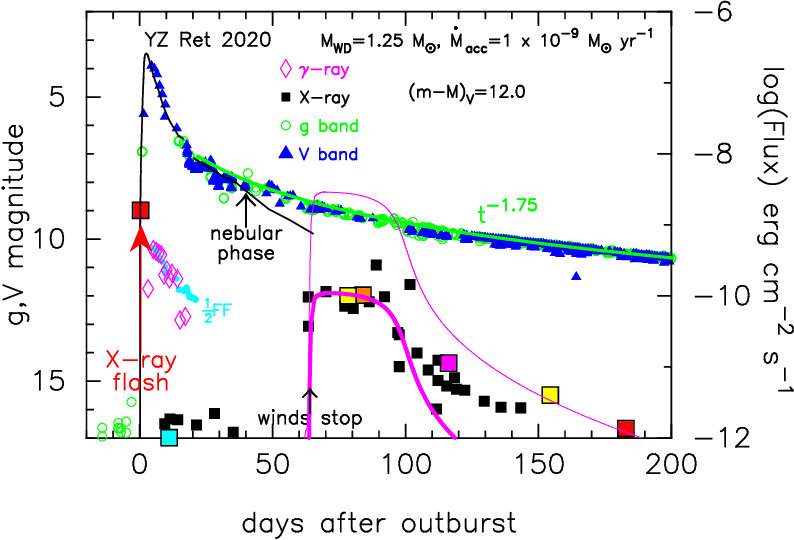}
\caption{
A summary of multiwavelength light curves of the YZ Ret 2020 outburst.
Optical $V$ and $g$ magnitudes, X-ray fluxes (filled black squares by Swift,
filled red squares by eROSITA, filled cyan square by NuSTAR,  
filled orange square by NICER, filled yellow squares by XMM-Newton,
filled magenta square by Chandra), 
gamma-ray fluxes (open magenta diamonds by Fermi/LAT)
are taken from Figure 2 of \citet{kon22wa}.
The origin of time is set to be $t_{\rm OB}= (t=0) =$ UT 2020 July 7.63 
$=$ JD 2,459,038.13 $=$ MJD 59,037.63.
Solid lines show
a 1.25 $M_\sun$ WD model with $\dot{M}_{\rm acc}=1\times 10^{-9} ~M_\sun$
yr$^{-1}$ on to the WD \citep{kat26hsa}.
The thick green line labeled $t^{-1.75}$ shows a light curve model
in the nebular phase taken from \citet{hac23k}.
The thick magenta line corresponds to the blackbody X-ray (0.3-10 keV) flux,
which is reduced by a factor of 26.3 than the original value 
(thin magenta line).
The cyan symbols labeled ``${1 \over 2}$FF'' depict a slope half
as fast as that of free-free emission (a half decline slope of 
the $V$/$g$ light curve).
\label{optical_mass_yz_ret_full_cyc_x55z02o10ne03_no4}}
\end{figure*}

A classical nova is an unstable hydrogen burning of a hydrogen-rich
envelope on a mass-accreting white dwarf (WD) \citep[e.g.,][for recent fully
self-consistent nova outburst models]{kat22sha, kat26hsa}.
The classical nova YZ Ret went into outburst in 2020. 
We show a summary of multiwavelength light curves
of YZ Ret in Figure \ref{optical_mass_yz_ret_full_cyc_x55z02o10ne03_no4}, 
where we regard the origin of time as the outburst day, 
$t_{\rm OB}= t =0=$ UT 2020 July 7.63 $=$ JD 2459038.13 
$=$ MJD 59037.63. 

Soft X-rays were detected with the eROSITA instrument on board 
Spectrum-Roentgen-Gamma (SRG) on UT 2020 July 7 (day 0) 
before optical brightening \citep{kon22wa}.
This is the first detection of an X-ray
flash phase in classical novae \citep{kon22wa, kat22shb}.
GeV gamma rays were observed during $\sim 10$ days in the post-maximum phase,
which indicates a strong shock \citep{sok22ll}.
Shocks play an essential role in gamma-ray emission \citep[see, e.g.,][for
a review]{cho21ms}.  Hard X-rays are detected from day $\sim 10$
\citep{sok22ll}.  The supersoft X-ray source (SSS) phase starts on day 63
and lasts until day $\sim 100$.

YZ Ret was also observed with longer wavelengths such as optical,
infrared, and radio \citep[e.g.,][]{sok22ll}.
\citet{hac23k} presented nova outburst models based on the 
post-maximum nova wind evolution \citep{kat94h} and clarified
the $V$ light curve evolution of YZ Ret as well as the gamma-ray and
X-ray light curves.  

Very recently, \citet{kat25hsa, kat26hsa} calculated
fully self-consistent nova outburst cycles and presented their $V$ light
curves based on free-free emission of nova winds \citep{hac06kb}. 
In the present paper, using these new models calculated by
\citet{kat26hsa}, we reanalyze the multiwavelength light curves
of YZ Ret.


This paper is organized as follows. First we apply our self-consistent 
nova outburst models to the $V$/$g$ light curves of YZ Ret and 
determine the WD mass and mass accretion rate on to the WD
in Section \ref{self-consistent_nova_models}.
Emissions from a shocked shell are summarized separately in Section
\ref{shock_nova_ejecta}.  
Discussion and conclusions follow in Sections \ref{discussion} and
\ref{conclusions}, respectively.

\begin{figure*}
\epsscale{1.15}
\plottwo{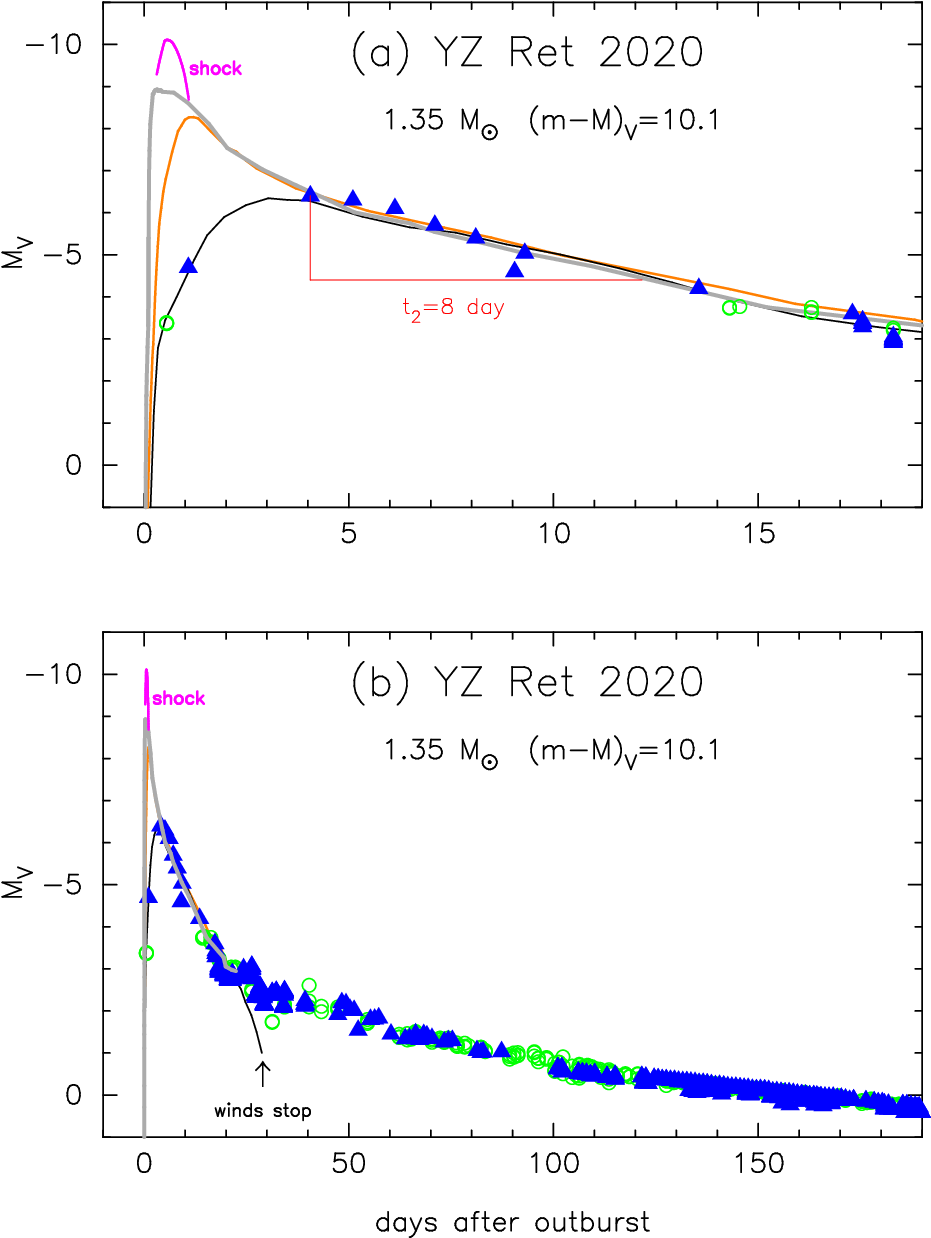}{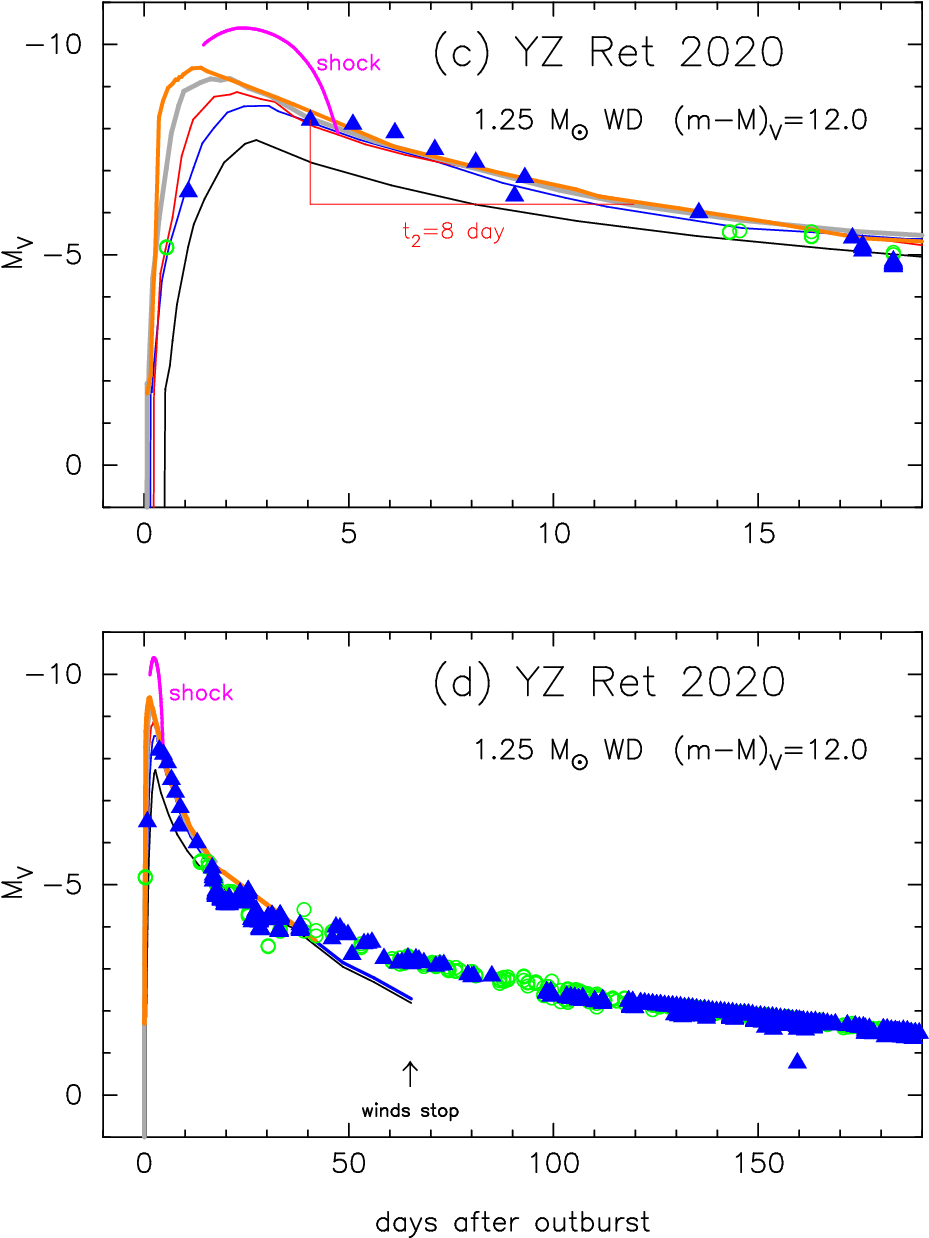}
\caption{
Optical $V$/$g$ light curve models of the YZ Ret 2020 outburst.
Optical $V$ (filled blue triangles) and $g$ (open green circles) magnitudes
are taken from Figure 2 of \citet{kon22wa}.
(a)(b) The 1.35 $M_\sun$ WD light curve models
with three values of the mass-accretion rate, i.e.,
$1\times 10^{-11}$ (thick gray line),
$5\times 10^{-10}$ (orange line), and
$5\times 10^{-9}$ (black line) $M_\sun$ yr$^{-1}$ \citep{kat26hsa},
are shown for $(m-M)_V=10.1$.
The magenta line labeled shock denotes a $V$ light curve for an optically
thick shocked shell in V1674 Her \citep[$M_{\rm WD}=1.35 ~M_\sun$,
$\dot{M}_{\rm acc}=1\times 10^{-11} ~M_\sun$ yr$^{-1}$;][]{hac26kv1674her3}.
(c)(d) The 1.25 $M_\sun$ WD models with five values of the mass-accretion
rate, i.e.,
$5\times 10^{-11}$ (thick orange line),
$1\times 10^{-10}$ (thick gray line),
$5\times 10^{-10}$ (red line),
$1\times 10^{-9}$ (blue line), and
$5\times 10^{-9}$ (black line) $M_\sun$ yr$^{-1}$ \citep{kat26hsa},
are shown for $(m-M)_V=12.0$.
The magenta line labeled shock denotes a $V$ light curve for an optically
thick shocked shell in V1500 Cyg \citep[$M_{\rm WD}=1.25 ~M_\sun$,
$\dot{M}_{\rm acc}=5\times 10^{-11} ~M_\sun$ yr$^{-1}$;][]{hac26kv1674her3}.
\label{optical_mass_yz_ret_x55z02o10ne03_no2}}
\end{figure*}

\section{Fully self-consistent nova models}
\label{self-consistent_nova_models}

\citet{kat25hsa, kat26hsa} calculated nova outburst cycles for a 1.35 
$M_\sun$ WD with three mass accretion rates of $1\times 10^{-11}$, 
$5\times 10^{-10}$, and $5\times 10^{-9} ~M_\sun$ yr$^{-1}$ and also
for a 1.25 $M_\sun$ WD with five mass accretion rates of
$5\times 10^{-11}$, $1\times 10^{-10}$, $5\times 10^{-10}$,
$1\times 10^{-9}$, and $5\times 10^{-9} ~M_\sun$ yr$^{-1}$
as tabulated in Table 1 of \citet{kat26hsa}.
They used their own Henyey type evolution code consistently combined with
steady state wind mass loss solutions as a surface boundary condition,
and calculated the WD structures from the center of the WD up to the
WD photosphere.
The method of their numerical calculation is explained
in \citet{kat22sha, kat24M1213, kat25hsa, kat26hsa}.

\subsection{Optical $V$/$g$ light curve}
\label{optical_light_curve}

Based on the free-free emission model of nova winds \citep{hac06kb},
\citet{kat25hsa, kat26hsa} presented $V$ light curves calculated with 
\begin{equation}
L_{V, \rm ff,wind} = A_{\rm ff} ~{{\dot M^2_{\rm wind}} 
\over{v^2_{\rm ph} R_{\rm ph}}},
\label{free-free_flux_v-band}
\end{equation}
where the coefficient $A_{\rm ff}$ was determined by 
\citet{hac10k, hac15k, hac16k} and \citet{hac20skhs} for various 
sets of WD mass and chemical composition, $\dot{M}_{\rm wind}$ is
the wind mass-loss rate, $v_{\rm ph}$ is the wind velocity at the
WD photosphere, and $R_{\rm ph}$ is the photospheric radius of the WD. 
\citet{kat25hsa, kat26hsa} compared them with the V1674 Her, KT Eri,
V339 Del, V597 Pup, and SMC Nova 2016-10a light curves, and
\citet{hac26ksbright, hac26kv1723sco} applied them
to V1500 Cyg 1975, V1674 Her 2021,
CP Lac 1936, CP Pup 1942, V838 Her 1991, V597 Pup 2007, V5583 Sgr 2009\#3,
V5589 Sgr 2012\#1, and V1723 Sco 2024,
and obtain the WD masses, mass-accretion rates, distance moduli
in the $V$ band, $(m-M)_V$, of each nova.
They reasonably reproduced the optical $V$ light curves of each nova
and confirmed a wide applicability of these light curve models to novae.
Therefore, we apply these fully self-consistent nova outburst models to YZ Ret.
In what follows, we fit these model $V$ light curves with YZ Ret,
and estimate the WD mass, mass-accretion rate on to the WD,
distance modulus in the $V$ band, $(m-M)_V$, and so on.

Panels (a) and (b) of Figure \ref{optical_mass_yz_ret_x55z02o10ne03_no2}
show 1.35 $M_\sun$ WD light curve models
with the three mass-accretion rates of
$1\times 10^{-11}$ (thick gray line),
$5\times 10^{-10}$ (orange line), and
$5\times 10^{-9}$ (black line) $M_\sun$ yr$^{-1}$ \citep{kat26hsa}.  With
the distance modulus in the $V$ band of $(m-M)_V=10.1$, we convert the
apparent $V$ magnitude of observation into the absolute $V$ magnitude, i.e.,
\begin{equation}
M_V = m_V - (m-M)_V,
\label{covert_absolute_mag}
\end{equation}
where $M_V$ is the absolute $V$ magnitude, $m_V$ is the apparent $V$
magnitude, and $(m-M)_V$ is the distance modulus in the $V$ band of each nova.
The light curves are calculated with Equation (\ref{free-free_flux_v-band})
based on the free-free emission light curve model and the largest
$\dot{M}_{\rm acc}$ model reproduces well the rising and decay phases.
We adopt the model of $\dot{M}_{\rm acc}=5\times 10^{-9} ~M_\sun$ yr$^{-1}$
(black line) among the three models.
We have examined the best fit distance modulus in the $V$ band
to be $(m-M)_V=10.1\pm 0.2$ by the least square fit with the observed 
$V$ data. Hereafter, we call this Model A.

Panels (c) and (d) of Figure \ref{optical_mass_yz_ret_x55z02o10ne03_no2}
show 1.25 $M_\sun$ WD models with the five mass-accretion rates of
$5\times 10^{-11}$ (thick orange line),
$1\times 10^{-10}$ (thick gray line),
$5\times 10^{-10}$ (red line),
$1\times 10^{-9}$ (blue line), and
$5\times 10^{-9}$ (black line) $M_\sun$ yr$^{-1}$.
Here, we adopt $(m-M)_V=12.0$.
We select the blue line of $\dot{M}_{\rm acc}=1\times 10^{-9} ~M_\sun$
yr$^{-1}$ among the five model $V$ light curves.
We have examined the best fit distance modulus in the $V$ band
to be $(m-M)_V=12.0\pm 0.2$ by the least square fit with the observed 
$V$ data.  Hereafter, we call this Model B.

We have the two WD models which reasonably reproduce the early optical 
$V$/$g$ light curves of YZ Ret.  One is a 1.35 $M_\sun$ WD with
$\dot{M}_{\rm acc}=5\times 10^{-9} ~M_\sun$ yr$^{-1}$ (model A) and
the other is a 1.25 $M_\sun$ WD with $\dot{M}_{\rm acc}=1\times 10^{-9}
~M_\sun$ yr$^{-1}$ (model B).  In what follows, we resolve this degeneracy 
in light of the other information.

\subsection{distance and reddening}
\label{distance_reddening}
We adopt the distance of $d=2.53 \pm 0.26$ kpc from
the Gaia data release 3 (DR3) \citep{bai21rf}.
The Galactic coordinates of YZ Ret is 
$(\ell, b)= (265\fdg 39744, -46\fdg 39540)$.
YZ Ret is located at the height from the Galactic plane
$z= -1.8$ kpc and belongs to the thick disk component of our Galaxy
\citep{sok22ll}.
Assuming that the reddening toward YZ Ret saturates,
we take the reddening of $E(B-V)=0.0152\pm 0.0009$
from the 2D Galactic dust extinction map of \citet{schlaf11f}.

The distance modulus in the $V$ band toward YZ Ret is calculated to be 
\begin{equation}
(m-M)_V=3.1 E(B-V) + 5 \log \left({{d} \over {\rm 10 ~pc}}\right)
=12.06\pm0.2.
\label{distance_reddening_law}
\end{equation}
This is consistent with $(m-M)_V= 12.0\pm0.2$ of model B (1.25 $M_\sun$ WD
with $\dot{M}_{\rm acc}=1\times 10^{-9} ~M_\sun$ yr$^{-1}$).


\begin{figure*}
\epsscale{1.0}
\plottwo{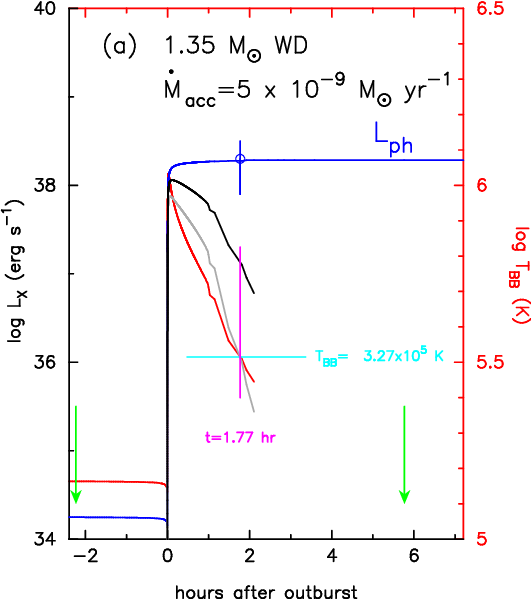}{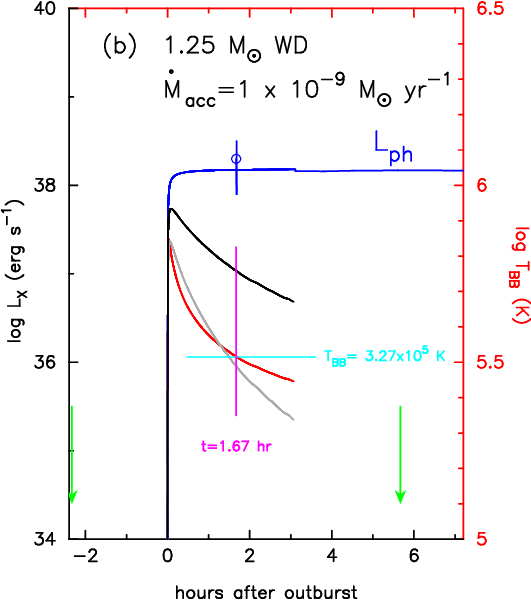}
\caption{
Our model X-ray luminosities (black and gray lines) and photospheric
temperature (red line) against time in units of hours.
The black and gray lines denote 0.2-10 keV and 0.3-10 keV X-ray bands,
respectively.  These lines end when optically thick winds
emerge from the photosphere, because X-rays are absorbed by winds.
We also add the photospheric luminosity ($L_{\rm ph}$) 
by the blue line labeled $L_{\rm ph}$.  
The red line crosses the temperature of $T_{\rm BB}=3.27\times 10^5$ K
at 1.77 hr in panel (a) Model A: 1.35 $M_\sun$ WD with 
$\dot{M}_{\rm acc}= 5\times 10^{-9} ~M_\sun$ yr$^{-1}$
or at 1.67 hr in panel (b) Model B: 1.25 $M_\sun$ WD with 
$\dot{M}_{\rm acc}= 1\times 10^{-9} ~M_\sun$ yr$^{-1}$,
which corresponds to the time of eROSITA detection.
X-rays were not detected at the downward green arrows, 4 hr before/after
the epoch of detection.
Both Model A and B are
consistent with the X-ray detection/non-detection
with eROSITA \citep{kon22wa}. 
\label{x-ray_flash_yz_ret_1.25_1e9}}
\end{figure*}

\subsection{X-ray flash phase}
\label{x-ray_flash_phase}

A brief soft X-ray bright phase was observed with the SRG/eROSITA on day 0
before optical brightening \citep{kon22wa}.  With the assumption of
spherically symmetric emissions and the Gaia eDR3 distance,
\citet{kon22wa} obtained the photospheric luminosity
of $L_{\rm ph}=(2.0 \pm 1.2) \times 10^{38}$ erg s$^{-1}$ and
the photospheric temperature of $kT_{\rm ph}=28.2^{+0.9}_{-2.8}$ eV 
from their 36 s observation.

\citet{kat22shb} presented models of X-ray flash
and argued that the duration of $< 8$ hr and the blackbody
temperature of $kT_{\rm ph}\approx 28$ eV of this X-ray flash 
are reproduced in massive WDs ($M_{\rm WD} \gtrsim 1.3~M_\sun$)
with a mass-accretion rate of 
$\dot M_{\rm acc} \lesssim 5 \times 10^{-9}~M_\sun$ yr$^{-1}$.
They excluded the less massive WDs of 1.0 and  $1.2~M_\sun$, 
but not examined $1.25~M_\sun$ WDs.

This very short X-ray flash phase corresponds to the very early phase
of a nova outburst 
as in Figure \ref{optical_mass_yz_ret_full_cyc_x55z02o10ne03_no4}.  
Figure \ref{x-ray_flash_yz_ret_1.25_1e9} shows our model calculations 
for this X-ray flash phase.  
Panel (a) shows model A (1.35 $~M_\sun$ WD with $\dot{M}_{\rm acc}=
5\times 10^{-9} ~M_\sun$ yr$^{-1}$)
while panel (b) depicts model B (1.25 $~M_\sun$ WD with 
$\dot{M}_{\rm acc}= 1\times 10^{-9} ~M_\sun$ yr$^{-1}$).
\citet{kon22wa} obtained the blackbody temperature of 
$k T_{\rm ph}=28.2^{+0.9}_{-2.8}$ eV, which corresponds to
$T_{\rm BB}= (28.2^{+0.9}_{-2.8})\times$ 11604.45 K
$=3.27^{+0.1}_{-0.3}\times 10^5$ K.
The horizontal cyan line of $T_{\rm BB}=3.27\times 10^5$ K crosses
the red line, of model A at $t=1.77$ hr in panel (a),
or of model B at 1.67 hr in panel (b).

No soft X-rays were observed 4 hr before/after the detection
\citep{kon22wa} as indicated by the downward green arrows.
Both model A and B are consistent with these no detections of X-rays.
The photospheric luminosity (open blue circle with error bars)
is also consistent with $L_{\rm ph}$ (blue line) for the both models.  
Thus, both the models are consistent with the X-ray flash observation. 

\subsection{Supersoft X-ray source phase}
\label{sss_phase}

YZ Ret entered the supersoft X-ray source (SSS) phase on day $\sim$63
as shown in Figure \ref{optical_mass_yz_ret_full_cyc_x55z02o10ne03_no4}.  
In model A ($1.35 ~M_\sun$ WD), optically thick winds stop on day $\sim$29
(Figure \ref{optical_mass_yz_ret_x55z02o10ne03_no2}(b))
and the nova enters the SSS phase.  This epoch (day 29) is too early
to be compatible with the X-ray observation (day $\sim 63$).
In our model B ($1.25 ~M_\sun$ WD), winds stop on day $\sim 65$
(Figure \ref{optical_mass_yz_ret_x55z02o10ne03_no2}(d)),
which is roughly consistent with the X-ray observation as shown
in Figure \ref{optical_mass_yz_ret_full_cyc_x55z02o10ne03_no4}.

Our supersoft X-ray model light curves are shown in 
Figure \ref{optical_mass_yz_ret_full_cyc_x55z02o10ne03_no4},
where the original model flux (thin magenta line) is much larger 
than the observation.  We suppose that about 96\% of the X-rays 
are occulted by the elevated disk edge \citep{hac23k}.
This occulted X-ray light curve (thick magenta line) is consistent 
with the observation during day 65-100 \citep[see also, e.g.,][for
other examples]{nes13oh}.  Thus, the X-ray flux of model B is consistent
with the X-ray observation rather than that of model A.

For the occultaion of X-ray, we assume that 
an L$_1$ stream with an increased mass
accretion rate impacts the disk edge and makes a splay
when the companion has been
irradiated by the very hot and luminous WD \citep[see, e.g., 
][for details]{sch97mm, hac25kw}.
After the elevation of the disk edge gradually lowered after day 110
(after the end of irradiation),
its flux approaches the original flux of the thin magenta line  
\citep[see ][for more details]{hac23k}.

To summarize, our model B satisfies all the requirements of
the $V$/$g$ light curve, $(m-M)_V$, X-ray flash and SSS phases while
our model A shows disagreements with the $(m-M)_V$ and SSS phase.
Therefore, we adopt model B as the best-fit model among 
\citet{kat26hsa}'s fully self-consistent nova outburst models.

\section{Shock in nova ejecta}
\label{shock_nova_ejecta}

GeV gamma-rays and hard X-rays were detected in some classical novae.
GeV gamma-rays are observed in the post-maximum phase for a few tens of days
\citep[e.g.,][]{abd10, ack14aa, li17mc, gor21ap}.
Hard X-rays are observed in an intermediate phase of a nova outburst
\citep[e.g.,][]{llo92ob, bal98ko, muk01i}.
It has been argued that both GeV gamma-rays and hard X-rays originate
from a strong shock \citep[see, e.g., ][for a review]{cho21ms}.

Based on \citet{kat22sha}'s fully self-consistent nova explosion models,
\citet{hac22k} showed that a strong shock naturally arises
outside the WD photosphere in the post-maximum phase.
In their models, before the optical maximum,
later ejected matter has a smaller velocity than that of earlier ejecta
and therefore the ejecta expands while, after the optical maximum,
later ejected matter has a larger velocity than that of earlier ejecta
so that it catches up with the earlier ejecta and makes a strong shock.
Thus, a shock naturally arises after the maximum expansion
of the WD photosphere ($=$optical maximum for the free-free emission model
light curves) and propagates far outside the WD photosphere.
This shock formation mechanism reasonably explains gamma-ray
and hard X-ray emissions in classical novae \citep[e.g.,
YZ Ret, V339 Del, and V392 Per in][respectively]{hac23k, hac24km,
hac25kv392per}.

\subsection{Shock formation}
\label{shock_formation}

A strong shock naturally arises outside the WD photosphere
just after the optical maximum \citep{hac22k}.
Our model $V$ light curve of 
a 1.25 $~M_\sun$ WD with $\dot{M}_{\rm acc}= 1\times 10^{-9} 
~M_\sun$ yr$^{-1}$ (model B) attains its peak on day 2.3 
(blue line in Figure \ref{optical_mass_yz_ret_x55z02o10ne03_no2}(c)).
On the other hand, the Fermi/LAT detected GeV gamma-rays from YZ Ret
on MJD 59040.49 \citep{sok22ll}, that is, on day 2.86.
This is consistent with the requirement that the optical $V$ maximum
precedes the detection of GeV gamma-rays.
Therefore, we adopt the optical $V$ maximum on day 2.3 (model B) and
a shock arises soon after the optical $V$ maximum for YZ Ret.

\subsection{Multiple velocity systems in the ejecta}
\label{multiple_velocities_ejecta}

\citet{mcl42} classified multiple velocity systems of
absorption and emission lines in an early phase of a nova into 
three categories, i.e., pre-maximum, principal, and diffuse enhanced
absorption/emission line systems.
\citet{hac22k} interpreted these velocity systems
as illustrated in Figure 2 of \citet{hac23k} based on their
shock formation mechanism: the pre-maximum system is the velocity
of earliest wind from the nova, the principal system is the velocity
of the shocked shell, and the diffuse enhanced system is the
velocity of the inner (latest) wind from the nova. 

\citet{sok22ll} presented two P Cygni velocity components, $-1200$ and
$-2700$ km s$^{-1}$ \citep[][]{ayd20bc} 
on JD 2459046.67, i.e., on day 8.54,
about 6 days after the optical maximum.
We assign these two velocities to the principal and
diffuse enhanced systems, i.e., $v_{\rm shock}=v_{\rm p}=
1200$ km s$^{-1}$ and $v_{\rm wind}=v_{\rm d}= 2700$ km s$^{-1}$, respectively.
Here, $v_{\rm shock}$ is the velocity of the shocked shell,
$v_{\rm p}$ is the velocity of the principal system, 
$v_{\rm wind}$ is the velocity of the inner (latest) wind,
and $v_{\rm d}$ is that of the diffuse-enhanced system.
Both $v_{\rm p}$ and $v_{\rm d}$ are the same as those
in \citet{hac23k} and our results derived from $v_{\rm p}$ and $v_{\rm d}$
are the same as those in \citet{hac23k}.

\subsection{Shock temperature}
\label{shock_temperature}

The temperature at the reverse shock is given by
\begin{eqnarray}
k T_{\rm sh}& \sim & {3 \over 16} \mu m_p 
\left( v_{\rm wind} - v_{\rm shock} \right)^2 \cr
& \approx & 1.0 {\rm ~keV~} 
\left( {{v_{\rm wind} - v_{\rm shock}} \over  
{1000 {\rm ~km~s}^{-1}}} \right)^2,
\label{shock_kev_energy}
\end{eqnarray}
where 
$T_{\rm sh}$ is the temperature at the shock
\citep[see, e.g.,][]{met14hv}, 
$\mu$ is the mean molecular weight ($\mu =0.5$ for hydrogen plasma),
and $m_p$ is the proton mass.
Substituting $v_{\rm shock}= v_{\rm p}=1200$ km s$^{-1}$ and 
$v_{\rm wind}= v_{\rm d}=2700$ km s$^{-1}$ into Equation 
(\ref{shock_kev_energy}),
we obtain the same post-shock temperature as in \citet{hac23k}, i.e.,
$k T_{\rm sh}\sim 2.3$ keV several days after optical maximum.

\subsection{Shock luminosity}
\label{shock_luminosity}

The reverse shock converts mechanical energy of the wind into thermal energy
of the gas as \citep{met14hv}
\begin{eqnarray}
L_{\rm sh}& \sim & {{9}\over {32}} {\dot M}_{\rm wind} 
{{( v_{\rm wind} - v_{\rm shock} )^3} \over {v_{\rm wind}}} \cr
&=& 1.8\times 10^{37}{\rm ~erg~s}^{-1}
\left( {{{\dot M}_{\rm wind}} \over 
{10^{-4} ~M_\sun {\rm ~yr}^{-1}}} \right) \cr
 &  & \times
\left( {{{v_{\rm wind} - v_{\rm shock}} \over {1000{\rm ~km~s}^{-1}}}}
\right)^3
\left( {{{1000{\rm ~km~s}^{-1}} \over {v_{\rm wind}}} }\right). 
\label{shocked_energy_generation}
\end{eqnarray}
Substituting $\dot{M}_{\rm wind}= 1.4 \times 10^{-4} ~M_\sun$ yr$^{-1}$
on day 3.0 (model B: a 1.25 $M_\sun$ WD with $\dot{M}_{\rm acc}=
1\times 10^{-9} ~M_\sun$ yr$^{-1}$), we obtain the post-shock energy of
$L_{\rm sh} \sim 4\times 10^{37}$ erg s$^{-1}$.

\citet{sok22ll} obtained the flux of GeV gamma-ray (0.1--300 GeV) to be 
$L_{\gamma}= 3\times 10^{35}$ erg s$^{-1}$ and the optical luminosity
to be $L_{\rm opt}= 7\times 10^{38}$ erg s$^{-1}$ on day 4.6.
The gamma-to-optical ratio is $L_{\gamma}/L_{\rm opt} = 4\times 10^{-4}$.
The ratio of $L_{\gamma}/L_{\rm sh}= 0.007$,
about 0.7\% conversion rate, is consistent with
\begin{equation}
\epsilon_{\rm nth} \epsilon_{\gamma} = L_{\gamma}/ L_{\rm sh}
\lesssim 0.03,
\label{gamma_efficiency}
\end{equation}
where $\epsilon_{\rm nth}\lesssim 0.1$ is the fraction of the shocked
thermal energy to accelerate nonthermal particles, and
$\epsilon_{\gamma}\lesssim 0.1$ is the fraction
of this energy radiated in the Fermi/LAT band \citep[typically 
$\epsilon_{\rm nth} \epsilon_{\gamma} < 0.03$;][]{met15fv}.

\subsection{Decay of GeV gamma-ray luminosity}
\label{decay_gamma-ray}

\citet{hac23k} approximately calculated the gamma-ray light curves
in their Figure 3, which are based on the relation,
\begin{equation}
m_{\gamma, \rm sh}(t) = {1\over 2} m_{V,\rm ff}(t) + {\rm ~constant}, 
\label{gamma_ray_half_magnitude}
\end{equation}
where $m_{V,\rm ff}(t)$ and $m_{\gamma, \rm sh}(t)$ are the $V$ 
magnitude of free-free emission luminosity and the magnitude of
the Fermi/LAT band gamma-ray flux, respectively, as a function of time $t$,
which simply means that the decay trend (slope) of the gamma-ray flux
is about a half of the optical $V$ flux.  
This is because, in our model, the optical flux is dominated by free-free
emission and is given by $L_{V, \rm ff, wind}\propto (\dot{M}_{\rm wind}/
v_{\rm ph})^2 \propto f(t)$ from Equation (\ref{free-free_flux_v-band})
while the gamma-ray flux is given by $L_\gamma = \epsilon_{\rm nth}
\epsilon_\gamma L_{\rm sh} \propto 
(\dot{M}_{\rm wind}/v_{\rm wind})\propto [f(t)]^{1/2}$  from Equations 
(\ref{shocked_energy_generation}) and (\ref{gamma_efficiency}),
where $f(t)$ is a function of time $t$.  Here, we assume that
$v_{\rm wind} - v_{\rm shock}$ is almost constant near/around
the post-maximum phase.

We have calculated the gamma-ray flux (cyan dots labeled ``${1 \over 2}$FF''
in Figure \ref{optical_mass_yz_ret_full_cyc_x55z02o10ne03_no4})
from Equation (\ref{gamma_ray_half_magnitude}).
Its decay trend follows well that of the observed gamma-ray flux
(open magenta diamonds).

\subsection{Column density of ejecta}
\label{column_density_ejecta}

The wind mass loss rate substantially drops in the nebular phase, 
so that the velocity and mass of the shocked shell hardly changes.
Then, the column density of hydrogen is calculated from
$M_{\rm shell}= 4 \pi R_{\rm sh}^2 \rho h_{\rm shell}$,
where $\rho$ is the density in the shocked shell, 
and $h_{\rm shell}$ is the thickness of the shocked shell.
With the velocity of the shell $v_{\rm sh}= v_{\rm shell}= v_{\rm shock}
= v_{\rm p} = 1200$ km s$^{-1}$ and the shocked shell radius of 
$R_{\rm sh}(t)= v_{\rm shock}\times t$, we estimate the column density by
\begin{eqnarray}
N_{\rm H} & = & {{X \over m_p} {{ M_{\rm shell} }
\over {4 \pi R^2_{\rm sh}}}} \cr
 & \approx & 4.8\times 10^{22} {\rm ~cm}^{-2}
\left({X \over {0.5}}\right)
\left( {{M_{\rm shell}} \over {10^{-5} M_\sun}} \right)
\left( {{R_{\rm sh}} \over {10^{14} {\rm ~cm}}} \right)^{-2}
\cr
 & \approx & 6.4 \times 10^{20} {\rm ~cm}^{-2}
\left({X \over {0.5}}\right)
\left( {{M_{\rm shell}} \over {10^{-5} M_\sun}} \right) \cr
& & \times
\left( {{v_{\rm shell}} \over {1000 {\rm ~km~s}^{-1}}} \right)^{-2}
\left( {{t} \over {100~{\rm days}}} \right)^{-2}.
\label{column_density_hydrogen_time}
\end{eqnarray}
This gives $N_{\rm H}\approx 1.5\times 10^{20}$ cm$^{-2}$ for
$M_{\rm shell}= 3\times 10^{-6} ~M_\sun$ (model B: a 1.25 $M_\sun$ WD
with $\dot{M}_{\rm acc}=1\times 10^{-9} ~M_\sun$ yr$^{-1}$), 
$v_{\rm shell}=1200$ km s$^{-1}$, and $t=90$ days, which is the end
of the SSS phase.
  

\citet{pei20og} and \citet{ori22gg} analyzed the NICER data
(0.3--2 keV) on day 83 and 84 and obtained 
$T_{\rm ph}=$500,000--550,000 K and $N_{\rm H}= 3\times 10^{20}$ cm$^{-2}$.
This $N_{\rm H}$ is broadly consistent with our estimate
if we include the interstellar extinction $E(B-V)=0.015$, which is
converted to $N_{\rm H}= 6.86\times 10^{21} {\rm ~cm}^{-2} ~E(B-V)
= 1\times 10^{20}$ cm$^{-2}$ \citep{gue09o}.  The total $N_{\rm H}$
is $(1.5 + 1)\times 10^{20}$ cm$^{-2}$ on day 90. 

\begin{figure*}
\epsscale{0.95}
\plottwo{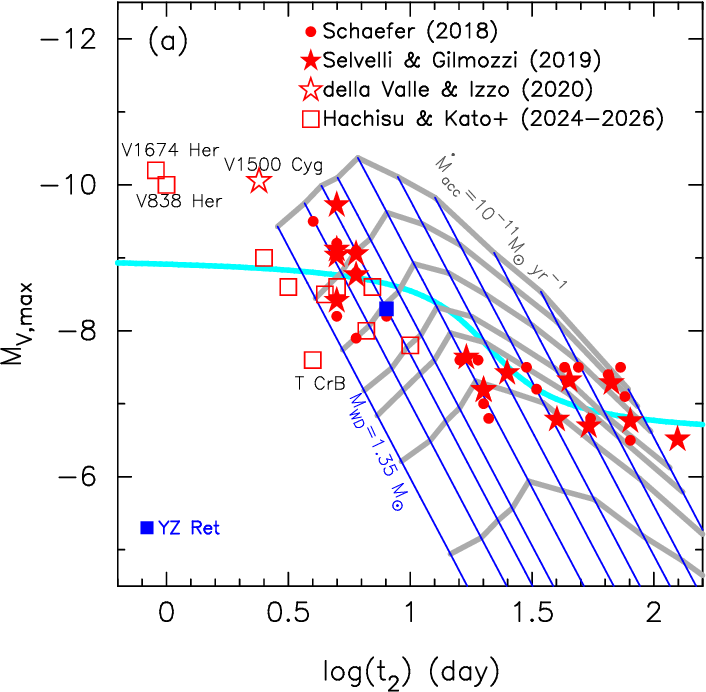}{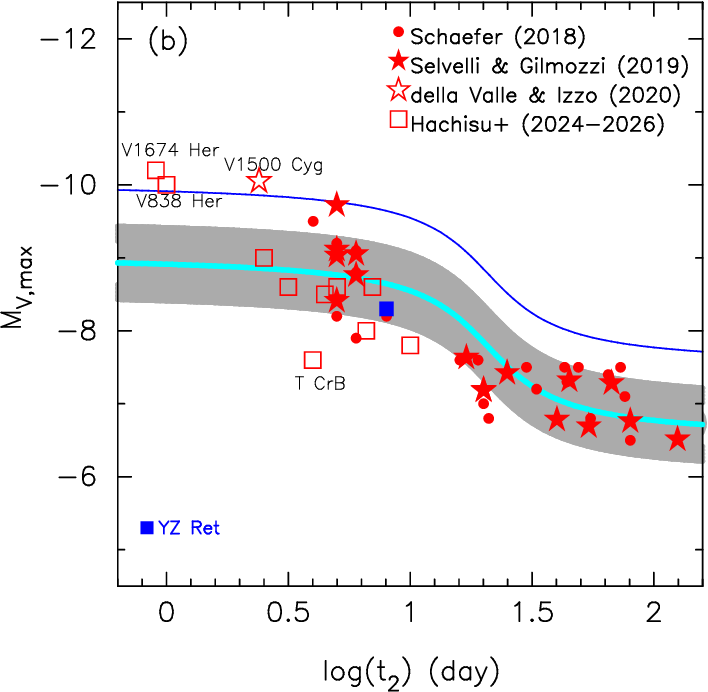}
\caption{
(a) MMRD diagram for classical novae.
The straight blue lines denote equi-WD mass lines, from left to right,
1.35, 1.3, 1.25, 1.2, 1.1, 1.0, 0.9, 0.8, 0.7, and $0.6~M_\sun$.
The thick solid gray lines depict equi-mass accretion rate lines, 
from lower to upper, $3\times 10^{-8}$,
$1\times 10^{-8}$, $5\times 10^{-9}$, $3\times 10^{-9}$, $1\times 10^{-9}$,
$1\times 10^{-10}$, and $1\times 10^{-11} M_\sun$~yr$^{-1}$.
These lines are taken from \citet{hac20skhs}.
The brightnesses of novae are calculated from free-free emission
luminosity of Equation (\ref{free-free_flux_v-band}).
Each filled red circle is taken from ``Golden sample'' of
\citet{schaefer18}, each filled star is from \citet{sel19},
and open star is from \citet{del20i}.
The 9 novae (open red squares; KT Eri, V339 Del, V392 Per, V1674 Her,
V838 Her, V597 Pup, V5583 Sgr, V5589 Sgr, and V1723 Sco) are
taken from \citet{hac25kw}, \citet{hac24km}, \citet{hac25kv392per},
\citet{hac26kv1674her3}, \citet{hac26ksbright}, and \citet{hac26kv1723sco}, 
respectively.  The thick solid cyan line indicates the empirical
line obtained by \citet{del20i}.
The position of YZ Ret (filled blue square) is taken from the present work.
We also add the position of T CrB for comparison.
(b) Same as panel (a), but we show only the position of each nova and
empirical MMRD line of \citet{del20i}.   The thick cyan line
indicates the same as the thick cyan line in panel (a), and light-gray
shadow line corresponds to its $\pm 0.5$ mag region.
The blue line is 1 mag above the thick cyan line.
If the peak absolute $V$ magnitude of a nova is brighter than
the blue line, it is dubbed a superbright nova \citep{del91}.
The three novae, V1500 Cyg, V838 Her, and V1674 Her are superbright novae
\citep{hac26ksbright}.
\label{max_t2_yz_ret_selvelli2019_schaefer2018_all_saio_kato2026}}
\end{figure*}

\subsection{Duration of shock}
\label{shock_duration}

A shock arises soon after optical maximum and disappears when the wind stops.
More accurately, the shock terminates when the latest wind reaches the shock
front \citep{hac22k}.
Our model predicts the shock duration by substituting $v_{\rm sh}\approx 
v_{\rm p}=1200$ km s$^{-1}$ (principal system) and
$v_{\rm ph}\approx v_{\rm d}=2700$ km s$^{-1}$ (diffuse-enhanced system),
and $t_{\rm ws}=63-3=60$ days (the wind duration
from the epoch when the shock arises) into Equation (3) in \citet{hac23k},
i.e.,
\begin{equation}
\tau_{\rm shock}= {{t_{\rm ws}} \over
{\left( 1- {{v_{\rm p}} \over {v_{\rm d}}}\right)}}.
\label{duration_of_shock}
\end{equation}
The shock duration is $\tau_{\rm shock}= 60/0.5555= 108$ days.
Therefore, we expect hard X-ray emission until day $108+3=111$.
This is broadly the same as in \citet{hac23k}.

\section{Discussion}
\label{discussion}

\subsection{pre-nova evolution}
\label{pre-nova_evolution}

\citet{schaefer22} obtained the pre-eruption orbital period of YZ Ret to be
$P_{\rm orb}= 0.1324539\pm 0.0000098$ day ($= 3.179$ hr) based on TESS
optical photometry.  YZ Ret was known as MGAB-V207
and classified as a nova-like VY Sculptoris variable by two fadings from
$V\sim 16$ to 17.2 and 18.0 mag \citep{mur19}.
The mass-accretion rate on to the WD was estimated to be
${\dot M}_{\rm acc} \gtrsim 2\times 10^{-9} ~M_\sun$ yr$^{-1}$ by
\citet{kat22shb} because of the suppression of dwarf nova outbursts
\citep[e.g.,][]{osa96}.
Both model A (1.35 $~M_\sun$ WD, ${\dot M}_{\rm acc}=5\times 10^{-9} 
~M_\sun$ yr$^{-1}$) and model B (1.25 $~M_\sun$ WD, ${\dot M}_{\rm acc}=
1\times 10^{-9} ~M_\sun$ yr$^{-1}$) broadly satisfy
this requirement of ${\dot M}_{\rm acc} \gtrsim 2\times 10^{-9}
~M_\sun$ yr$^{-1}$.

It should be noted that CP Lac is a classical nova that belongs to
the VY Sculptoris type \citep{honey98}.
\citet{peters06} obtained the orbital period of 0.145143(1) day ($=3.48$ hr),
which is close to that of YZ Ret ($3.179$ hr).  Both YZ Ret and
CP Lac belong to the VY Scl type and have a similar orbital period of
$P_{\rm orb}\sim$3.2-3.5 hr.

\citet{hac26ksbright} obtained a best-fit model of a 1.25 $~M_\sun$ WD
with $\dot{M}_{\rm acc}=1\times 10^{-9} ~M_\sun$ yr$^{-1}$ (model B)
for CP Lac.  This is the same model as that for YZ Ret.
The pre-nova evolutionary state is similar, that is, both are a twin.
These similarities support our results that both the WD mass 
(model B: 1.25 $M_\sun$) and mass accretion rate
on to the WD (model B: $1\times 10^{-9} ~M_\sun$ yr$^{-1}$) 
are common between YZ Ret and CP Lac.

\subsection{Position in the MMRD diagram}
\label{position_mmrd_diagram}

The maximum $V$ magnitude versus rate of decline (MMRD) diagram has been
used to discuss nova properties \citep[e.g., ][]{del20i}.
\citet{hac20skhs} presented theoretical MMRD
diagrams for the relations among various nova parameters, based on 
our database of theoretical light curves of free-free emission
(Equation (\ref{free-free_flux_v-band})) 
for various sets of WD mass and mass-accretion rate.
Figure \ref{max_t2_yz_ret_selvelli2019_schaefer2018_all_saio_kato2026}(a)
shows such an MMRD diagram.
The gray and blue lines indicate the equi-$\dot{M}_{\rm acc}$ and
equi-$M_{\rm WD}$ lines, respectively. 
For comparison, we also plot a popular MMRD diagram in Figure
\ref{max_t2_yz_ret_selvelli2019_schaefer2018_all_saio_kato2026}(b).

We have measured the $t_2$ time of YZ Ret to be $t_2=8$ day from
the $V$ data of \citet{kon22wa}'s extended data table
(filled blue triangles) as depicted in
Figure \ref{optical_mass_yz_ret_x55z02o10ne03_no2}.
Here $t_2$ is the 2 mag decay time from the optical $V$ peak.
The maximum $V$ magnitude of YZ Ret is estimated to be
$M_{V,\rm max}= m_{V,\rm max} -(m-M)_V = 3.7 - 12.0 = -8.3$.

We plot the MMRD point of YZ Ret (filled blue square) in Figure
\ref{max_t2_yz_ret_selvelli2019_schaefer2018_all_saio_kato2026}.
YZ Ret is located on the blue line of 1.25 $M_\sun$ WD and
the gray line of $\dot{M}_{\rm acc}=1\times 10^{-9} ~M_\sun$ yr$^{-1}$
in Figure \ref{max_t2_yz_ret_selvelli2019_schaefer2018_all_saio_kato2026}(a).
These values are just the same as those of model B.
We may conclude that YZ Ret is a typical very fast nova 
\citep[$t_2 \le 10$ days, ][]{pay57}.



\section{Conclusions}
\label{conclusions}

We reanalyzed multiwavelength light curves of the classical nova YZ Ret
with our fully self-consistent nova outburst models.
Our main results are summarized as follows:\\

\noindent
\begin{enumerate} 
\item Among 1.35 $M_\sun$ WD models
with three mass accretion rates of $1\times 10^{-11}$, 
$5\times 10^{-10}$, and $5\times 10^{-9} ~M_\sun$ yr$^{-1}$ and 
1.25 $M_\sun$ WD with five mass accretion rates of
$5\times 10^{-11}$, $1\times 10^{-10}$, $5\times 10^{-10}$,
$1\times 10^{-9}$, and $5\times 10^{-9} ~M_\sun$ yr$^{-1}$,
our $V$/$g$ light curve fitting in the early phase of the outburst
suggests two models: model A: 1.35 $M_\sun$ WD with $\dot{M}_{\rm acc}=
5\times 10^{-9} ~M_\sun$ yr$^{-1}$ for $(m-M)_V=10.1\pm 0.2$
and model B: 1.25 $M_\sun$ WD with $\dot{M}_{\rm acc}=1\times 10^{-9}
~M_\sun$ yr$^{-1}$ for $(m-M)_V=12.0\pm 0.2$.
\item We adopt the distance of $d=2.53 \pm 0.26$ kpc from the Gaia DR3
parallax \citep{bai21rf} and the reddening of $E(B-V)=0.0152\pm 0.0009$
from the 2D Galactic dust extinction map \citep{schlaf11f}.
Then, the distance modulus in the $V$ band is calculated to be
$(m-M)_V= 12.06\pm0.2$.
This suggests that model B
is consistent with
the distance modulus in the $V$ band of $(m-M)_V= 12.06\pm0.2$.
%
%
%
\item An X-ray flash phase was observed in YZ Ret with the SRG/eROSITA
\citep{kon22wa}.  Both models A and B satisfy the observational requirements
of this X-ray flash, i.e., the blackbody photospheric temperature,
photospheric luminosity, and very short duration.  
\item Models A and B reach optical maximum ($=$maximum expansion of the
WD photosphere) on day 3.0 and 2.3, respectively.
The first GeV gamma-rays were detected on day 2.86 \citep{sok22ll}.
Because GeV gamma-rays are emitted by a shock, only model B satisfies
the requirement that the optical maximum precedes the detection of
GeV gamma-rays \citep{hac22k}.
\item Models A and B stop winds and enter the supersoft X-ray source phase
on day 29 and 65, respectively.
The observed soft X-ray flux rapidly increased on day 63.
Therefore, the SSS phase in model A is too early to be compatible with,
but model B is consistent with, the X-ray observation.  
\end{enumerate}
Our model B (1.25 $M_\sun$ WD with $\dot{M}_{\rm acc}=1\times 10^{-9}
~M_\sun$ yr$^{-1}$) satisfies all the requirements of the observation.

\begin{acknowledgments}
We thank the anonymous referee for her/his detailed and
useful comments, which greatly improved the manuscript.
\end{acknowledgments}

\end{document}